\documentclass[driverfallback=dvipdfmx]{akari2017}

\shorttitle{AGN selection by SED fitting}
\shortauthors{T.-C. Huang et al.}

\begin{document}

\title{AGN selection by 18-band SED fitting in mid-infrared in the AKARI NEP deep field } 



\correspondingauthor{Ting-Chi Huang}
\email{s104022505@m104.nthu.edu.tw}

\author{Ting-Chi Huang} 
\affiliation{National Tsing hua University, No. 101, Section 2, Kuang-Fu Road, Hsinchu, Taiwan 30013}
\author{Tomotsugu Goto} 
\affiliation{National Tsing hua University, No. 101, Section 2, Kuang-Fu Road, Hsinchu, Taiwan 30013}
\author{Tetsuya Hashimoto} 
\affiliation{National Tsing hua University, No. 101, Section 2, Kuang-Fu Road, Hsinchu, Taiwan 30013}
\author{Nagisa Oi} 
\affiliation{Tokyo University of Science, 1-3 Kagurazaka, Shinjuku-ku, Tokyo 162-8601, Japan}
\author{Hideo Matsuhara} 
\affiliation{Institute of Space and Astronautical Science, Japan Aerospace Exploration Agency, 3-1-1 Yoshinodai, Chuo, Sagamihara, Kanagawa 252-5210, Japan}



\begin{abstract}
In this research, we provide a new, efficient method to select infrared (IR) active galatic nucleus (AGN). In the past, AGN selection in IR had been established by many studies using  color-color diagrams. However, those methods have a problem in common that the number of bands is limited. The AKARI North Ecliptic Pole (NEP) survey was carried out by the AKARI Infrared Camera (IRC), which has 9 filters in mid-IR with a continuous wavelength coverage from 2 to 24$\mu$m$^{-1}$. Based on the intrinsic different mid-IR features of AGN and star-forming galaxies (SFGs), we performed SED fitting to separate these two populations by the best-fitting model. In the X-ray AGN sample, our method by SED fitting selects 50$\%$ AGNs, while the previous method by colour criteria recovers only 30$\%$ of them, which is a significant improvement. Furthermore, in the whole NEP deep sample, SED fitting selects two times more AGNs than the color selection. This may imply that the black hole accretion history could be more stronger than people expected before. 
\end{abstract}



\setcounter{page}{1}





\section{Introduction}
Active Galactic Nucleus (AGN) plays an important role in galaxy evolution. It has been widely known that AGN's powerful energy comes from the accretion of its supermassive black hole (SMBH). Therefore, we can probe how galaxies and SMBHs have evolved with cosmic time by investigating AGNs. As a result, it is fundamental to have a efficient way to select AGNs. However, the X-ray and ultraviolet (UV) light emitted from the accretion disk are tend to be obscured by the dust and gas of AGN torus \citep[e.g.,][]{Alexander et al. 2001,Richard et al. 2003,Webster et al. 1995}. Under this situation, we fortunately can still detect AGN in IR from the thermal emission of dust. But here comes another problem that not only AGNs, but also star-forming galaxies (SFGs) can be observed in IR due to the polycyclic aromatic hydrocarbon (PAH) emission features at 3.3, 6.2, 7.7, 8.6 and 11.3 $\mu$m$^{-1}$. Hence, to select AGNs in IR, the separation between AGNs and SFGs is extremely essential and important.  
\\
Since AGN has a power-law like spectrum from the dust emission in mid-IR, previous studies frequently used some criteria in color-color diagrams to select AGNs \citep[e.g.,][]{Jarrett et al. 2011,Lacy et al. 2004,Richard et al. 2006,Stern et al. 2005}. However, selections by colors have a limitation of the number of bands. Futhermore, the observatories used in previous work generally have a detection gap in their wavelength range. For example, Spitzer IRAC has 4 filter at 3.6, 4.5, 5.8 and 8.0 $\mu$m$^{-1}$, but the next closest filter is the MIPS24 at 24 $\mu$m$^{-1}$ . AKARI Infrared Camera (IRC) is equipped with 9 filters in mid-IR, which continuously covers the wavelength range from 2 to 24 $\mu$m$^{-1}$  \citep{Matsuhara et al. 2006}. The AKARI IRC 9 bands in mid-IR allow us to separate AGNs and SFGs by SED fitting. In this work, we combined the mid-IR photometry from AKARI, WISE and Spitzer and performed SED fitting with 25 empirical templates to select AGNs by 18 mid-IR bands.  

\section{Data and Analysis}
\label{DandA}
We used the catalogue from the AKARI North Ecliptic Pole (NEP) deep survey \citep{Murata et al. 2013} and required all the objects having a well calculated magnitude at the AKARI 18 $\mu$m$^{-1}$ band. In total, we have 5761 objects. In addition to AKARI, we also included the data from GALEX, CFHT, WISE, Spitzer, and Herschel for SED fitting, but only a part of objects have been detected by each instruments other than AKARI. The redshift information of every NEP deep objects were calculated in \cite{Oi et al. 2014}. 
\\
We performed SED fitting by LEPHARE code to select AGNs. In the SED fitting, the UV and optical data are used to fit with stellar components \citep{Coleman et al. 1980, Kinney et al. 1996}, and the mid-IR data are used to fit with the models of AGNs and SFGs \citep{Polletta et al. 2007}. After the fitting, some criteria were requested to eliminate badly-fit results in our sample. We have 4833 objects in the final sample and we name it "the SED sample". To separate stars and galaxies, we followed the suggestion in \cite{Oi et al. 2014} using the CLASS$\_$STAR parameter from the CFHT source extraction software, SExtractor, of the CFHT z'-band. 

\section{Results}
\subsection{AGN selection}
\begin{figure}[!ht]
	\plotone{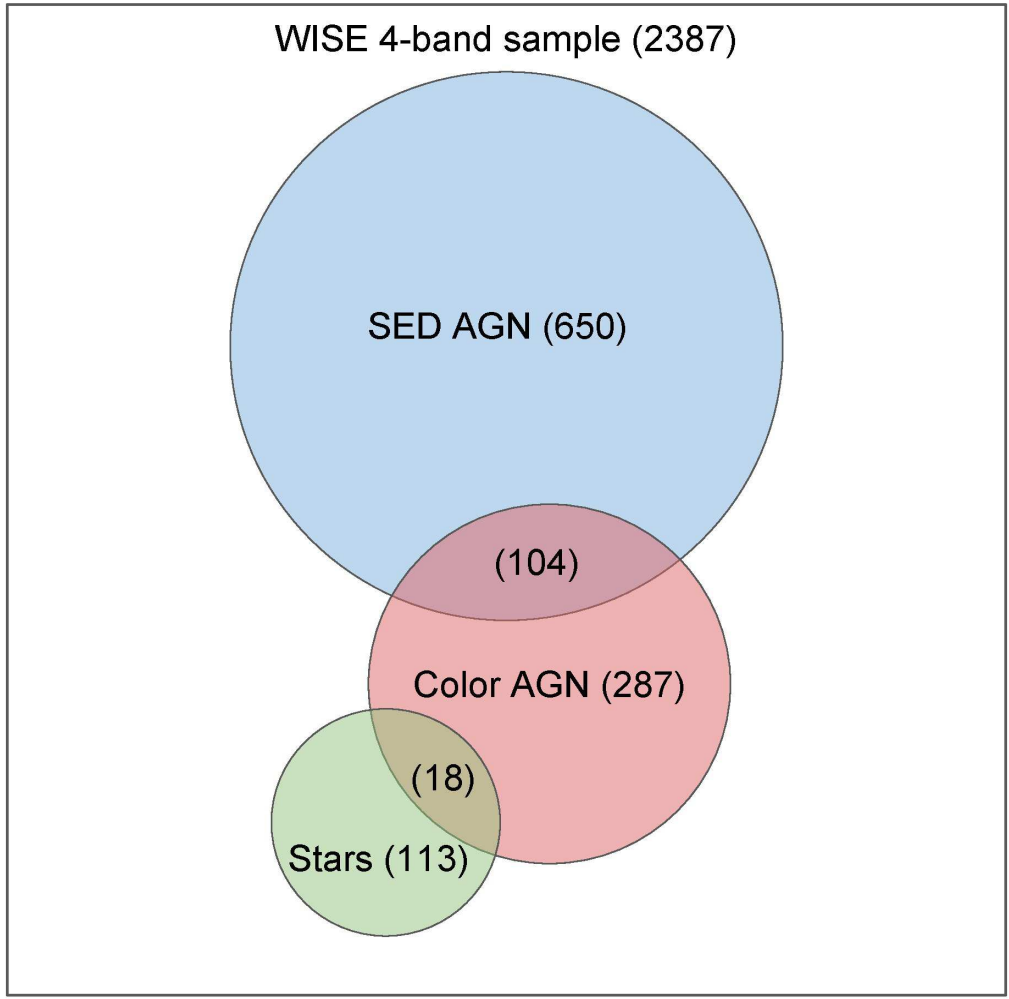}
\caption{The Venn diagram which shows the AGN selection by our SED fitting and the WISE color-color diagram. The numbers in parentheses mean the counts of the objects.}
\label{fig:Venn}
\end{figure}

We selected AGN in the galaxy sample by checking the best-fitting model. We used the SWIRE templates \citep{Polletta et al. 2007} to select AGN. The SWIRE templates contain 25 models, and 9 of them are AGN models, which include models of Seyfert 1.8, Seyfert 2, QSO2, QSO1, BQSO1, TQSO1, Mrk231, IRAS19254 and Torus. The other 16  SFGs models contain 3 elliptical galaxies, 7 spiral galaxies and 6 starbursts. Fig.~\ref{fig:Venn} shows the summary of AGN selection results from SED fitting and the WISE color-color diagram \citep{Wright et al. 2010,Jarrett et al. 2011}. In the subsample that all the objects have WISE 4-band magnitudes, our SED fitting selects nearly 2 times more AGNs than the criteria in color-color diagram. This result impies that AGN luminosity function and the cosmic black hole accretion history may be much stronger than people expected in the past.   

\subsection{AGNs missed by color selection}
\begin{figure}[!ht]
	\plotone{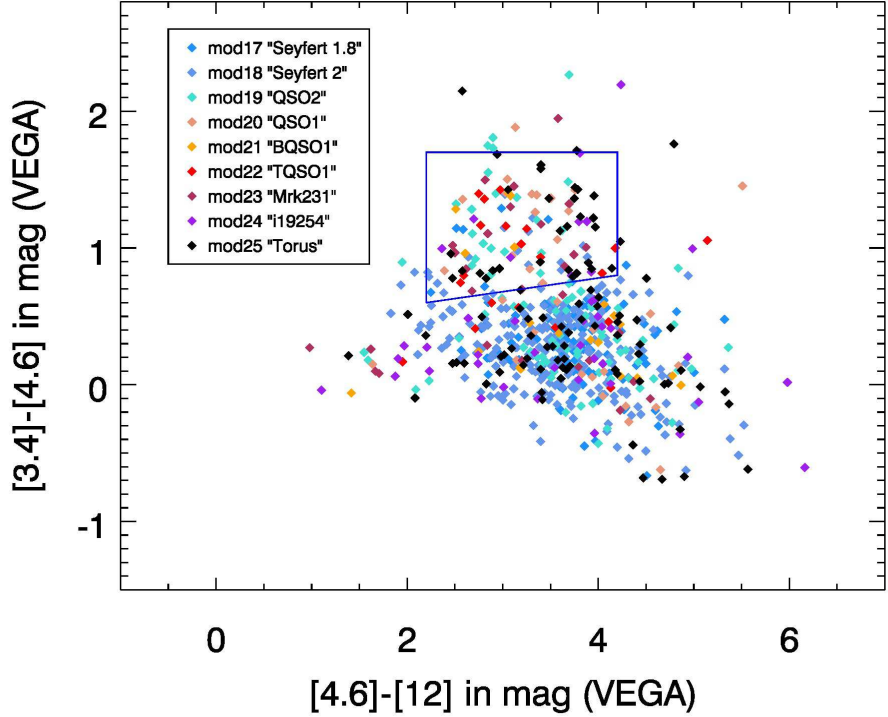}
\caption{The WISE color-color diagram with AGNs selected by our SED fitting. Different models are plotted in different colors as shown in the legend.}
\label{fig:ccdiagram_sedagnmod}
\end{figure}

There are many AGNs selected by the SED fitting but not by the color criteria. We show those SED AGNs in the WISE color-color diagram in Fig.~\ref{fig:ccdiagram_sedagnmod}. The color box is defined to select QSOs, so not surprisingly, there are many Seyfert 1.8 and Seyfert 2 AGN outside of the box. AGNs of these two models are moderately luminous or host-dominated AGNs. Therefore,  the result suggests that the WISE color box is not able to select AGNs completely. 

\subsection{Examination by X-ray AGN}
\begin{figure}[!ht]
	\plotone{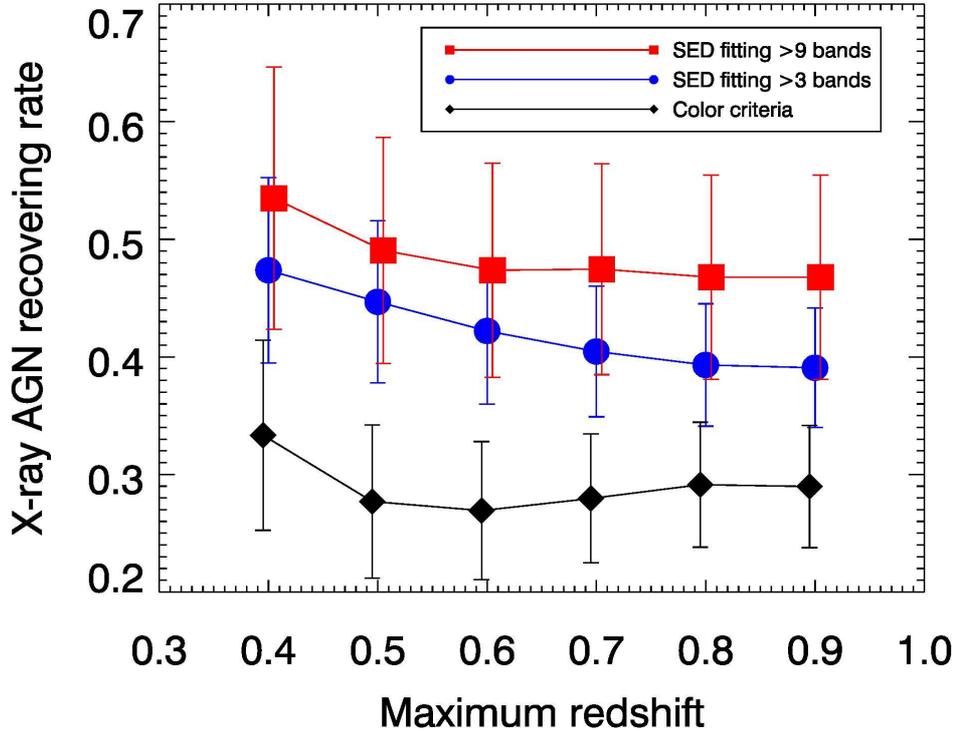}
\caption{The X-ray AGN recovering rate of the selection by our SED fitting and the color criteria. The red and blue lines are the results of SED fitting with different number of bands limitations. The black line shows the results from the selection by the WISE color-color diagram.}
\label{fig:recovering_rate}
\end{figure}

The X-ray catalogue of the AKARI NEP deep field from Chandra observation \citep{Krumpe et al. 2015} was included to compare the AGN selection by our SED fitting and the WISE color-color diagram. We matched the celestial coordinate to combine the X-ray catalogue with the NEP deep catalogue having WISE 4-band data in 1 arcsec tolerance radius. There are 254 X-ray objects in total and we regard all the X-ray sources as AGNs in this examination. In this X-ray sample, we compare the number fration of AGNs selected from the WISE color-color diagram and our SED fitting. The result shows that our SED fitting recovers more AGNs in the X-ray AGN sample (Fig.~\ref{fig:recovering_rate}). Moreover, if the number of band used in the SED fitting is larger, the recovering rate will become higher. Also, SED fitting can produce better results in low-redshift sample. 



\end{document}